\documentclass[%
 reprint,
 amsmath,amssymb,
 aps,
]{revtex4-2}

\usepackage{amsmath}
\usepackage{graphicx}
\usepackage{color}
\usepackage{dcolumn}
\usepackage{bm}
\usepackage[mathlines]{lineno}
\usepackage{array}
\usepackage{subfigure}
\usepackage{hyperref}
\hypersetup{hidelinks,
	colorlinks=true,
	allcolors=black,
	pdfstartview=Fit,
	breaklinks=true}
\usepackage{balance}

\begin{document}

\title{\boldmath Observation of the double Dalitz decay $\eta'\to e^+e^-e^+e^-$}


\author{
\begin{small}
\begin{center}
M.~Ablikim$^{1}$, M.~N.~Achasov$^{11,b}$, P.~Adlarson$^{70}$, M.~Albrecht$^{4}$, R.~Aliberti$^{31}$, A.~Amoroso$^{69A,69C}$, M.~R.~An$^{35}$, Q.~An$^{66,53}$, X.~H.~Bai$^{61}$, Y.~Bai$^{52}$
% ...
\\
\vspace{0.2cm}
(BESIII Collaboration)\\
\vspace{0.2cm} {\it
$^{1}$ Institute of High Energy Physics, Beijing 100049, People's Republic of China\\
% ...
$^{4}$ Bochum Ruhr-University, D-44780 Bochum, Germany\\
% ...
$^{11}$ G.I. Budker Institute of Nuclear Physics SB RAS (BINP), Novosibirsk 630090, Russia\\
% ...
$^{31}$ Johannes Gutenberg University of Mainz, Johann-Joachim-Becher-Weg 45, D-55099 Mainz, Germany\\
% ...
$^{35}$ Liaoning Normal University, Dalian 116029, People's Republic of China\\
% ...
$^{52}$ Southeast University, Nanjing 211100, People's Republic of China\\
$^{53}$ State Key Laboratory of Particle Detection and Electronics, Beijing 100049, Hefei 230026, People's Republic of China\\
% ...
$^{61}$ University of Jinan, Jinan 250022, People's Republic of China\\
% ...
$^{66}$ University of Science and Technology of China, Hefei 230026, People's Republic of China\\
% ...
$^{69}$ University of Turin and INFN, (A)University of Turin, I-10125, Turin, Italy; (B)University of Eastern Piedmont, I-15121, Alessandria, Italy; (C)INFN, I-10125, Turin, Italy\\
$^{70}$ Uppsala University, Box 516, SE-75120 Uppsala, Sweden\\
% ...
$^{b}$ Also at the Novosibirsk State University, Novosibirsk, 630090, Russia\\
% ...
}\end{center}
\vspace{0.4cm}
\end{small}
}


\begin{abstract}
Based on $(10087 \pm 44)\times10^6$  $J/\psi$ events collected with the BESIII detector at BEPCII, the double Dalitz decay $\eta'\to e^+e^-e^+e^-$ is observed for the first time via the $J/\psi\to\gamma\eta'$ decay process. The significance is found to be 5.7$\sigma$ with systematic uncertainties taken into consideration. Its branching fraction is determined to be $\mathcal{B}(\eta'\to e^+ e^- e^+ e^-) = (4.5\pm1.0(\mathrm{stat.})\pm0.5(\mathrm{sys.})) \times 10^{-6}$.

\end{abstract}

\maketitle


\section{\label{sec:level1}Introduction}
The double Dalitz decays $\mathcal{P}\to \ell^+ \ell^- \ell'^+\ell'^-$,
where $\mathcal{P}$ is a pseudoscalar meson ($\mathcal{P}=\pi^0,
\eta$, or $\eta'$) while $\ell$ and $\ell'$ are leptons ($\ell, \ell'=e,\mu$), are
expected to proceed through an intermediate state of two virtual
photons. These processes are of great interest for understanding the
pseudoscalar transition form factor (TFF) and the interactions between
pseudoscalars and virtual photons~\cite{2018REscr,2010TPetri}. These TFFs are necessary inputs to calculate the pseduoscalar-meson-pole contributions to the hadronic light-by-light scattering, which causes the second largest uncertainty in the Standard Model determination of the muon anomalous magnetic moment~\cite{g-2white, g-2, 2014Rafel, 2001Johan}.
Particularly, the double Dalitz decays of pseudoscalar mesons help to determine the TFFs in the small timelike momentum region, i.e. $m_{ll}^2\le q^2\le m_P^2$, with $m_{ll}$ the invariant mass of the dilepton and $m_P$ the mass of the pseudoscalar meson, and thus are suitable
to determine the slope of the TFFs at $q^2=0$~\cite{g-2white}.

The pseudoscalar meson double Dalitz decays have been elucidated by
many physics models, such as the hidden gauge model~\cite{2010TPetri},
the modified vector meson dominance model~\cite{2010TPetri}, the data driven approach~\cite{2018REscr},
and the resonance chiral symmetric
approach~\cite{2017HCzyz}. Experimental searches have been carried out
in a variety of processes. The $\pi^0$ Dalitz decay $\pi^0\to e^+ e^-
e^+ e^-$ has been intensively studied, and its branching fraction (BF)
is determined to be $(3.38\pm0.16)\times10^{-5}$~\cite{pdg}, which is
dominated by a measurement from the KTeV
Collaboration~\cite{KTeV}. The KLOE Collaboration reported the first
observation of $\eta\to e^+ e^- e^+ e^-$ with a BF of
$(2.4\pm0.2(\mathrm{stat.})\pm0.1(\mathrm{sys.}))\times10^{-5}$~\cite{KLOE}. The
WASA Collaboration set the upper limits at the 90$\%$ confidence level
of $\mathcal{B}(\eta\to e^+ e^- \mu^+ \mu^-)<1.6\times10^{-4}$ and
$\mathcal{B}(\eta\to \mu^+ \mu^- \mu^+ \mu^-) < 3.6 \times
10^{-4}$~\cite{WASA}. However, the double Dalitz decay of $\eta'$ has not been observed to date.

The theoretical predictions for $\eta'\to e^+ e^- e^+ e^-$ BF are in the range of $(2.1-2.4)\times
10^{-6}$~\cite{2018REscr, 2010TPetri}, which makes it possible to be observed based on
$(10087\pm44)\times10^6$ $J/\psi$ events collected in the years of
2009, 2012, 2018, and 2019 at BESIII~\cite{Njpsi} through the
$J/\psi\to\gamma\eta'$ process.

\section{\label{sec:level2}BESIII Detector and Monte Carlo Simulation}

The BESIII detector~\cite{Ablikim:2009aa} records symmetric $e^+e^-$ collisions
provided by the BEPCII storage ring~\cite{Yu:IPAC2016-TUYA01}, which operates
in the center-of-mass energy range between $2.0$ and $4.95$~GeV.
BESIII has collected large data samples in this energy region~\cite{Ablikim:2019hff}. The cylindrical core of the BESIII detector covers $93\%$ of the full solid angle and consists of a helium-based
 multilayer drift chamber~(MDC), a plastic scintillator time-of-flight
system~(TOF), and a CsI(Tl) electromagnetic calorimeter~(EMC),
which are all enclosed in a superconducting solenoidal magnet
providing a 1.0~T (0.9~T in
2012) magnetic field. The solenoid is supported by an
octagonal flux-return yoke with resistive plate counter muon
identification modules interleaved with steel.
The charged-particle momentum resolution at $1~{\rm GeV}/c$ is
$0.5\%$, and the $dE/dx$ resolution is $6\%$ for electrons
from Bhabha scattering. The EMC measures photon energies with a
resolution of $2.5\%$ ($5\%$) at $1$~GeV in the barrel (end cap)
region. The time resolution in the TOF barrel region is 68~ps, while
that in the end cap region was initially 110~ps. The end cap TOF
system was upgraded in 2015 using multigap resistive plate chamber
technology, providing a time resolution of
60~ps~\cite{etof}.

Simulated data samples produced with {\sc
geant4}-based~\cite{geant4} Monte Carlo (MC) software, which
includes the geometric description of the BESIII detector and its
response, are used to determine detection efficiencies
and to estimate backgrounds. The simulation models the beam
energy spread and initial state radiation in the $e^+e^-$
annihilations with the generator {\sc
KKMC}~\cite{ref:kkmc}. The inclusive MC sample includes both the production of the $J/\psi$
resonance and the continuum processes incorporated in {\sc
KKMC}. The known decay modes are modeled with {\sc
evtgen}~\cite{ref:evtgen} using branching fractions taken from the Particle Data Group (PDG)~\cite{pdg}, and the remaining unknown charmonium decays
are modeled with {\sc lundcharm}~\cite{ref:lundcharm}. Final state radiation
from charged final state particles is incorporated using {\sc
photos}~\cite{photos}. The decay $J/\psi\to\gamma\eta'$ is generated with an angular distribution of $1 + \cos^2_{\theta_\gamma}$, while the decay $\eta'\to e^+e^-e^+e^-$ is generated with a specific model based on the resonance chiral symmetric theory with SU(3) breaking~\cite{2017HCzyz}.

\section{\label{sec:level3} Data Analysis}

Four charged particles with zero net charge are required to be within
the range of $|\!\cos \theta|<0.93$, where $\theta$ is defined with
respect to the $z$ axis, which is the symmetry axis of the MDC.  The
distance of closest approach to the interaction point (IP) must be
less than 10 cm along the $z$ direction, and less than 1 cm in the
plane transverse to the $z$ axis.  Particle identification~(PID) for charged tracks combines measurements of the energy deposited in the MDC~(d$E$/d$x$) and the flight time in the TOF to form likelihoods the electron $[\mathcal{L}(e)]$, pion $[\mathcal{L}(\pi)]$, and kaon $[\mathcal{L}(K)]$ hypotheses.
Tracks are identified as electrons when the the electron hypotheses satisfy $\mathcal{L}(e)>0.001$ and has the highest PID likelihood among the three
hypotheses, i.e. $\mathcal{L}(e)>\mathcal{L}(\pi)$ and
$\mathcal{L}(e)>\mathcal{L}(K)$. To reduce background from hadrons and muons, the
electron and positron candidates with higher momentum must satisfy
$E/cp > 0.8$, where $E$ is the deposited energy in the EMC and
$p$ is the momentum measured by the MDC~\cite{Ep1, Ep2, Ep3}.

Photon candidates are identified using showers in the EMC. The deposited energy of each shower must be more than $25$~MeV in the barrel region ($|\!\cos \theta|< 0.80$) and more than $50$~MeV in the end cap region ($0.86 <|\!\cos \theta|< 0.92$). To exclude showers that originate from charged tracks, the angle between the direction of each shower in the EMC and the direction of the closest extrapolated charged track must be greater than $10^{\circ}$. To suppress electronic noise and showers unrelated to the event, the difference between the EMC time and the event start time is required to be within $[0, 700]$ ns.
At least one photon satisfying these selection criteria is required in the final state.

\begin{figure}[!http]
 \centering
\includegraphics[width=0.5\textwidth]{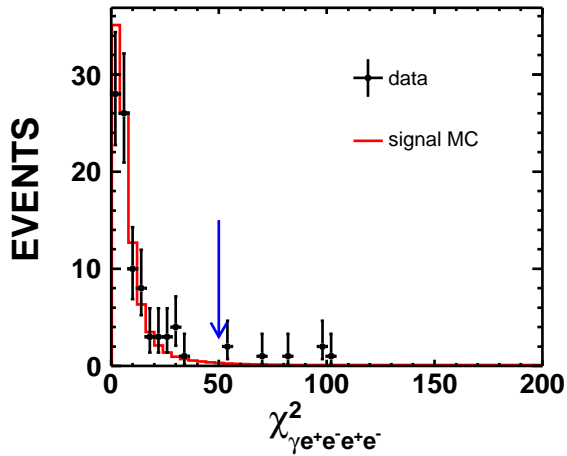}
\caption{The $\chi^2_{\gamma e^+e^-e^+e^-}$ distribution. The
  dots with error bars are from data, the red solid line is from signal
  MC. The arrow indicates the selection requirement of $\chi^2_{\gamma e^+e^-e^+e^-}<50$.}
\label{4C}
\end{figure}

A vertex constraint on the four electrons is imposed to ensure they
originate from the IP.  To improve the resolution and suppress the
background, a kinematic fit with an energy-momentum constraint (4C) is
performed under the hypothesis of $J/\psi \to \gamma e^+ e^- e^+
e^-$. For a small portion of events with multiple photon candidates, all of them are looped in the 4C fit. According to the simulation study, the $\chi^2_{\gamma e^+e^-e^+e^-}$ distribution of the 4C fit could identify the correct photon candidate effectively, so only the combination with the smallest $\chi^2_{\gamma e^+e^-e^+e^-}$ is retained.  The $\chi^2_{\gamma e^+e^-e^+e^-}$ distributions from data and signal MC are in a good agreement, as shown in Fig.~\ref{4C}. Events
with $\chi^2_{\gamma e^+e^-e^+e^-}<50$ are kept for further
analysis. To suppress background events with $\gamma \pi^+\pi^-e^+e^-$
in the final state, a 4C kinematic fit is performed under the
hypothesis of $J/\psi \to \gamma \pi^+\pi^-e^+ e^-$, and
$\chi^2_{\gamma e^+e^-e^+e^-}$ is required to be less than
$\chi^2_{\gamma \pi^+\pi^-e^+e^-}$.

\begin{figure}[!http]
 \centering
\includegraphics[width=0.5\textwidth]{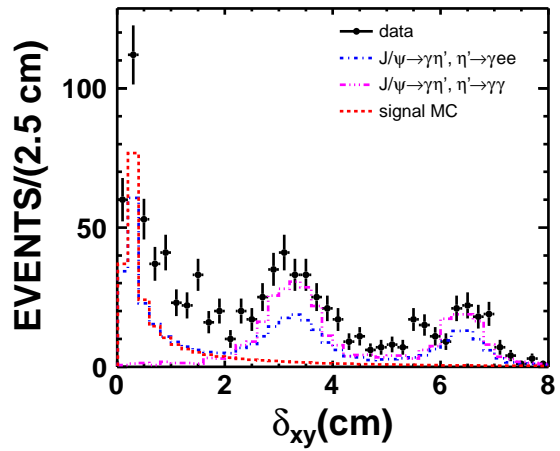}
\caption{The distribution of $\delta_{xy}$. The black dots with error bars represent data, the red dashed histogram shows the signal MC simulation. The blue dash-dotted and magenta dash-dot-doted histograms show the simulated $\gamma$ conversion samples from $J/\psi\to\gamma\eta'$, $\eta'\to\gamma e^+e^-$ and $J/\psi\to\gamma\eta'$, $\eta'\to\gamma \gamma$, respectively.}
\label{conv}
\end{figure}

The processes $J/\psi\to\gamma\eta'$ with $\eta'\to\gamma\gamma$,
$\eta'\to\gamma e^+e^-$ and $\eta'\to\gamma \rho^0$, $\rho^0\to
e^+e^-$ may contribute peaking backgrounds to the invariant mass
$M(e^+e^-e^+e^-)$ distributions, if the photons subsequently convert
into $e^+e^-$ pairs in the beam pipe or the inner wall of the MDC. To distinguish the $\gamma$ conversion events from
signal events, the Photon Conversion Finder package~\cite{PCF, yuanxq} is used. Two
variables are considered: (i) The distance $\delta_{xy}=\sqrt{R_{x}^2+R_{y}^2}$
from the reconstructed vertices of the $e^+e^-$ pairs to the IP in the
transverse plane. Here, $R_{x}$ and $R_{y}$ represent the coordinates
of the reconstructed vertex position in the $x$ and $y$ directions. The distributions of $\delta_{xy}$ for the data, $\gamma$ conversion background, and signal from MC simulation are shown in Fig.~\ref{conv}. (ii)
The angle $\theta_\mathrm{eg}$ between the momentum vector of the
radiative photon and the direction from IP to conversion point. In each event, there are four possible $e^+e^-$ pairs. The event with any $e^+e^-$ pair satisfying $\delta_{xy} \ge 2$ cm and $\cos\theta_\mathrm{eg}\ge-0.50$ is rejected. This $\gamma$ conversion veto requirement removes most of the $\gamma$ conversion backgrounds with a relative detection
efficiency loss less than $20\%$.

\begin{figure}[!http] 
 \centering
\includegraphics[width=0.5\textwidth]{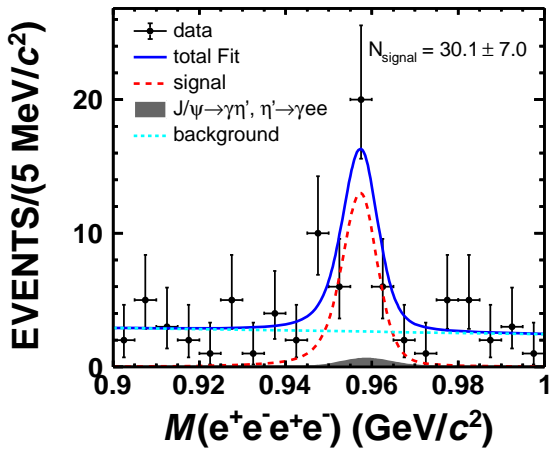} 
\caption{The $M(e^+
e^- e^+ e^-)$ distribution of data and the fitting results. The
  dots with error bars are data, the red dashed line is the signal
  shape, and the solid blue line is the total fit result. The gray area
  represents the peaking background from $J/\psi\to \gamma\eta',
  \eta'\to\gamma e^+e^-$, and the cyan dotted line is a linear
  function. }
\label{fit}
\end{figure}

The signal region is defined as $0.9 < M(e^+ e^- e^+ e^-) <1.0$ GeV/$c^2$. After applying the above selection criteria, the total detection efficiency is determined to be $(12.67\pm0.02)\%$ and the number of survived events is 86 as shown in the Fig.~\ref{fit}.

The possible background contaminations are studied with an
inclusive MC sample of $1.0011\times 10^{10}$ simulated events and data samples taken at $\sqrt s$ =  \mbox{$3.08$} and $\sqrt s$ =  \mbox{$3.773$ GeV}. Since the cross section of $e^{+}e^{-}\to\gamma\eta'$ is less than 1 nb~\cite{2017HCzyz}, the possible background from this process is negligible. For the
dominant background channels, dedicated exclusive MC samples are
generated to estimate their contributions. After applying all selection criteria, most
backgrounds are negligible. The only remaining peaking background in the $e^+ e^- e^+ e^-$
mass spectrum comes from
$J/\psi\to\gamma\eta'$, $\eta'\to\gamma e^+e^-$, with the number of
expected events being $2.48\pm0.18$ after taking into consideration the
BF from the PDG~\cite{pdg}.

After the above selection, the resulting $M(e^+ e^- e^+ e^-)$
distribution is shown in Fig.~\ref{fit}, where a clear $\eta'$ signal
is visible. An unbinned extended maximum likelihood fit is performed
to determine the $\eta'$ signal yield. The signal probability density
function (PDF) is represented by the signal MC shape convolved with a
Gaussian function to account for the resolution difference between
data and MC simulation. The shape for the nonpeaking background is
described by a linear function. The background yield and its PDF
parameters are allowed to float in the fit. The peaking background
from the $\gamma$ conversion process of $J/\psi\to\gamma\eta'$,
$\eta'\to\gamma e^+e^-$ is described by the MC-simulated shape~\cite{chuxk} with
the yield fixed at 2.48. 

The BF of $\eta'\to
e^+ e^- e^+ e^-$ is calculated as
\begin{equation}
\begin{aligned}
\mathcal{B}(\eta'\to e^+ e^- e^+ e^-) &=\frac{N_\mathrm{signal}}{N_\mathrm{J/\psi} \cdot \mathcal{B}(J/\psi \to \gamma \eta')  \cdot \epsilon}\\
&=(4.5\pm1.0) \times 10^{-6},
\end{aligned}
\end{equation}
where $N_\mathrm{signal}=30.1\pm7.0$ is the signal yield obtained from
fitting, $N_\mathrm{J/\psi}$ is the number of $J/\psi$ events,
$\epsilon$ is the MC-determined detection efficiency, and
$\mathcal{B}(J/\psi \to \gamma \eta')$ is the BF of $J/\psi \to \gamma
\eta'$ from the PDG~\cite{pdg}. The significance
is determined by evaluating
$\sqrt{-2\mathrm{ln}(\mathcal{L}_\mathrm{0}
  /\mathcal{L}_\mathrm{max})}$, where
$\mathcal{L}_\mathrm{0}$($\mathcal{L}_\mathrm{max}$) is the smeared
likelihood without (with maximum) signal events. When the systematic uncertainties are taken into account by convolving the likelihood function of the signal yield with a Gaussian function, the significance is estimated to  5.7$\sigma$

\section{\label{sec:level4}Systematic Uncertainties} Table~\ref{sys}
lists sources of systematic uncertainties considered in the BF
measurement.  Most systematic uncertainties are determined from
comparisons of low-background, high-statistics data samples and MC
simulation samples.

The PID and tracking efficiencies of electrons are determined from a
control sample of radiative Bhabha scattering $e^+e^- \to \gamma
e^+e^-$ (including $J/\psi \to \gamma e^+e^-$) corresponding to the
center-of-mass energy of the $J/\psi$ resonance. For the electron PID
study, the same PID requirements as used in this analysis are
applied to the control samples. Similarly, for the electron-tracking
study, the same conditions on the polar angle and the distances of
closest approach are used. Differences in PID (tracking)
efficiencies between the data and MC simulations are obtained for each
bin of a two-dimensional distribution of the momentum
(transverse momentum) versus the polar angle of the electron
tracks. These results are subsequently used to determine the overall
weighted differences per track for PID and tracking, and the obtained
PID and tracking uncertainties for electrons are $3.8\%$ and 2.2$\%$,
respectively. The photon detection efficiency is studied with the
control sample based on $J/\psi \to \pi^+\pi^-\pi^0$,
$\pi^0\to\gamma\gamma$ events~\cite{sys_pid_pi}. The difference
between data and MC simulation is $0.5\%$ ($1.5\%$) for a photon in
the EMC barrel (end cap) region. Since most photon candidates are
located in the barrel region, the difference $0.5\%$ is taken
as systematic uncertainty.

In the 4C kinematic fit, the track helix parameters are corrected to
reduce the differences between data and MC simulation~\cite{sys_4Cfit,
chuxk}. The $2.7\%$ difference between the efficiencies with and
without helix parameter corrections is taken as the systematic
uncertainty. The systematic uncertainty related with the $\gamma$
conversion veto criterion has been investigated with a control sample
of $e^+e^-\to\gamma e^+e^-$ at $\sqrt{s}=3.08$
GeV, where the invariant mass of $e^+e^-$ is limited to a range of extremely small mass ($< 10$ MeV/$c^2$). The relative difference of efficiency associated with the
$\gamma$ conversion rejection criterion between data and MC simulation
is $1.4\%$. Since the $\gamma$ conversion veto criterion is used for
two $e^+e^-$ pairs in the final state, the total systematic
uncertainty is $5.6\%$.

In the nominal signal MC model, the parameters describing $\eta-\eta'$
mixing and the $\pi^0\to\gamma\gamma$ decay width are fixed. To
estimate the systematic uncertainty due to the form-factor
parametrizations, signal MC events are also generated using the
global fit results from Ref.~\cite{2017HCzyz}. The relative difference
7.1$\%$ in the detection efficiency is taken as the uncertainty
associated with the signal model.

The uncertainty due to the nonpeaking background shape is estimated
by varying the PDF shape and fitting range. The fit range is changed
to $(0.89,1.01)$ or $(0.91, 0.99)$ GeV/$c^2$. A second-order
Chebyshev polynomial is selected as an alternative background shape. A
series of alternative fits are performed for all combinations of fit
ranges and nonpeaking background shapes to account for possible
correlations. The largest relative difference of the signal yield with
respect to the nominal values, 2.7$\%$, is taken as the systematic
uncertainty. In the nominal fit, the number of peaking background
events from $J/\psi \to \gamma \eta' , \eta' \to\gamma e^+ e^-$ is
fixed. To determine the systematic uncertainty, the number of peaking
background events is varied within its uncertainty, and the resulting
difference is negligible.

The uncertainty in the quoted BF of $J/\psi\to\gamma\eta'$ is
$1.3\%$~\cite{pdg}. The uncertainty in the number of $J/\psi$ events
is determined to be 0.4$\%$~\cite{Njpsi}.

Assuming that all systematic uncertainties in Table~\ref{sys} are
independent, the total systematic uncertainty, obtained from their
quadratic sum, is $10.8\%$.

\begin{table}[h] 
\caption{The sources of systematic uncertainties and their estimated magnitudes. The
negligible uncertainty is marked with an ellipsis.}  
\begin{center}
\renewcommand\arraystretch{1.3} 
\resizebox{\linewidth}{!}{
\begin{tabular}{l|c } 
\hline \hline 
Sources & Systematic uncertainties
(in $\%$)\\ 
\hline 
PID&$3.8$\\ 
MDC tracking&$2.2$ \\ 
Photon detection & 0.5\\ 
4C kinematic fit & 2.7\\ 
Veto of $\gamma$ conversion &5.6\\ 
Signal model &7.1 \\ 
Fit range and background shape &2.7\\ 
Peaking background & …\\ 
Quoted BF &$1.3$ \\ 
Number of $J/\psi$ &0.4\\ 
\hline 
Total uncertainty & 10.8\\ 
\hline \hline
\end{tabular}} 
\end{center} 
\label{sys} 
\end{table}

\section{\label{sec:level5}Summary} 
Using a sample of about $1\times10^{10}$
$J/\psi$ events collected with the BESIII detector, we observe
the double Dalitz decay process $\eta ' \rightarrow e^+e^-e^+e^-$ via
$J/\psi\to\gamma\eta'$, with a significance of 5.7$\sigma$ after
taking the systematic uncertainty into consideration. The BF of $\eta'
\rightarrow e^+e^-e^+e^-$ is measured to be
$(4.5\pm1.0(\mathrm{stat.})\pm0.5(\mathrm{sys.})) \times 10^{-6}$. The
measured BF is consistent with the theoretical predictions within the
uncertainties and provides new information for the studies about
$\eta'$ TFF and the interactions between $\eta'$ and virtual
photons~\cite{2018REscr, 2010TPetri}.

\begin{acknowledgments}
The BESIII collaboration thanks the staff of BEPCII and the IHEP computing center for their strong support. We thank Feng Xu for useful discussions. This work is supported in part by National Key R$\&$D Program of China under Contracts Nos. 2020YFA0406400, 2020YFA0406300; National Natural Science Foundation of China (NSFC) under Contracts Nos. 11625523, 11635010, 11735014, 11822506, 11835012, 11935015, 11935016, 11935018, 11961141012, 12022510, 12025502, 12035009, 12035013, 12061131003, 12165022; the Chinese Academy of Sciences (CAS) Large-Scale Scientific Facility Program; Joint Large-Scale Scientific Facility Funds of the NSFC and CAS under Contracts Nos. U1732263, U1832207; CAS Key Research Program of Frontier Sciences under Contract No. QYZDJ-SSW-SLH040; 100 Talents Program of CAS; INPAC and Shanghai Key Laboratory for Particle Physics and Cosmology; ERC under Contract No. 758462; European Union Horizon 2020 research and innovation programme under Contract No. Marie Sklodowska-Curie grant agreement No 894790; German Research Foundation DFG under Contracts Nos. 443159800, Collaborative Research Center CRC 1044, FOR 2359, GRK 214; Istituto Nazionale di Fisica Nucleare, Italy; Ministry of Development of Turkey under Contract No. DPT2006K-120470; National Science and Technology fund; Olle Engkvist Foundation under Contract No. 200-0605; STFC (United Kingdom); The Knut and Alice Wallenberg Foundation (Sweden) under Contract No. 2016.0157; The Royal Society, UK under Contracts Nos. DH140054, DH160214; The Swedish Research Council; U. S. Department of Energy under Contracts Nos. DE-FG02-05ER41374, DE-SC-0012069.

\end{acknowledgments}

%

\end{document}